\documentclass[prb,twocolumn,preprintnumbers,amsmath,amssymb,floatfix]{revtex4-1}
\usepackage[dvipsnames]{xcolor}
\usepackage{soul}
\sethlcolor{green}

\usepackage{graphicx}% Include figure files
\usepackage{dcolumn}% Align table columns on decimal point
\usepackage{bm,textcomp}% bold math
\usepackage{booktabs,threeparttable}
\usepackage{subfigure}
\usepackage{epstopdf}
\usepackage{color}
\makeatletter % the following 3 lines change section titles to lowercase
\def\@hangfrom@section#1#2#3{\@hangfrom{#1#2#3}}
\makeatother
\makeatletter % the following 4 lines make [] style for reference
\def\@biblabel#1{[#1]}
\makeatother

\newcommand{\cm}[1]{} % comment
\DeclareGraphicsExtensions{.eps,.mps,.pdf,.jpg,.png}
\definecolor{bg}{rgb}{0.75,.75,.67}
\begin{document}

\title{Atomistic origin of stress overshoots and serrations in a CuZr metallic glass}
%\date{\today}% It is always \today, today,%  but any date may be explicitly specified
%\email{chunguang.tang@unsw.edu.au or 0123.tang@gmail.com}
\author{Chunguang Tang$^{1*}$, Kevin Laws$^{1}$, Michael Ferry$^{1}$}
\address{$^1$ School of Materials Science and Engineering, The University of New South Wales, NSW 2052, Australia}

\begin{abstract}
In this work we use molecular dynamics simulations to study the stress overshoots of metallic glass Cu$_{50}$Zr$_{50}$ in three scenarios (unloading-reloading, slide-stop-slide, and stress serrations) that are associated with shear band relaxation. We found that, after the elastic recovery effect is factored out, atomic volume in the shear band barely changes during compressive relaxation but decreases during tensile relaxation, while local fivefold symmetry increases consistently for both cases. We propose that the atomistic mechanism for the related stress overshoots is due to the relaxation of structural symmetry, instead of free volume, in the shear band. Upon unloading, a propagating shear band continues for some time before arrested, which results in a stress undershoot and could contribute to material fatigue under cyclic elastic loads. We did not directly observe stress serrations via molecular dynamics simulations due to the very high simulated strain rates. While athermal quasistatic simulations produce serrated flow stress, we note that such serrations result from global avalanches of shear events rather than the relaxation of the shear band. %\hlt{Our studies provide atomistic insights on shear-banding dynamics and deepen the understanding of inhomogeneous mechanical response of metallic glasses.}
\end{abstract}

\maketitle
\section{Introduction} 

In the supercooled liquid region above their glass transition temperature ($T_g$), monolithic bulk metallic glasses deform homogeneously and exhibit stress overshoots which are more pronounced at lower temperatures and higher strain rates \cite{kawamura_deformation_1996, kawamura_stress_1997}. Below $T_g$, however, the plastic deformation of metallic glasses is highly localized in the form of shear bands and stress overshoots are generally not expected. Nevertheless, under several special conditions, stress overshoots have also been observed. First, during incremental cyclic nanoindentations of a series of BMGs \cite{yang_strain_2006, tekaya_quasi-static_2009, pan_deformation-induced_2010, lashgari_effect_2016} the load is transiently higher than the normal trend shortly after a full reloading.  Second, in slide-stop-slide compression tests \cite{maass_shear-band_2012,daub_effective_2014} a stress overshoot is observed when the sample, which is already in a steady flow stress state, is reloaded from a fixed-strain relaxation or aging. Third, the stress strain curves of monotonic uniaxial compressions at enough low strain rates exhibit flow serration \cite{dalla_torre_negative_2006, song_flow_2008, song_capturing_2010, maass_propagation_2011, maass_shear-band_2015}, which results from alternating stick and slip of a major shear band or several shear bands \cite{sun_plasticity_2010}. Such flow serrations are more pronounced at higher temperatures and lower strain rates \cite{maass_shear-band_2015}, in contrast to those \cite{kawamura_deformation_1996, kawamura_stress_1997} in the supercooled liquid region.

The shear-band-related overshoots below Tg are explained mainly by two categories of models based on the free volume concept or the shear transformation zone (STZ) theory. For cyclic nanoindentation, the overshoot was attributed to the net annihilation of free volume during the unloading process \cite{yang_strain_2006}. However, a series of simulation studies indicate that atoms are prone to shearing with either positive or negative free volume \cite{srolovitz_atomistic_1983,yang_structures_2016,tang_formation_2016}.  According to the STZ theory \cite{falk_dynamics_1998} STZs have two distinct orientations (denoted as `$+$' and `$-$') with the `$+$' state being saturated with plastic strain and the `$-$' state being able to be sheared. A steady stress state is reached when the creation and destruction of STZs and the rearrangement between the two orientations are balanced. The density of STZs depends on a state variable called effective temperature. By introducing the thermal contribution to the relaxation of effective temperature, some authors \cite{daub_effective_2014} found the number of STZs decreases during the aging stage of the slide-stop-slide experiments, which in turn causes a stress overshoot.

%(cite Zaccone, Origin of stress overshoot in amorphous solids
So far, a fundamental atomistic-scale understanding is still missing for the overshoot. Moreover, a general explanation for the three overshoot scenarios has not been established although they are interlinked: the cyclic loading case can transit into the slide-stop-slide case if the unloading rate and, hence, the unloading amount becomes infinitely small, and the latter merges into the monotonic loading situation as the stop time decreases to zero.  By performing atomistic simulations of tensile and compressive deformation of metallic glass Cu$_{50}$Zr$_{50}$, we report herein the atomistic-scale structural and topological features that generally explain the stress overshoot.

\begin{figure*}[t!]
\centering
\includegraphics[width=6in]{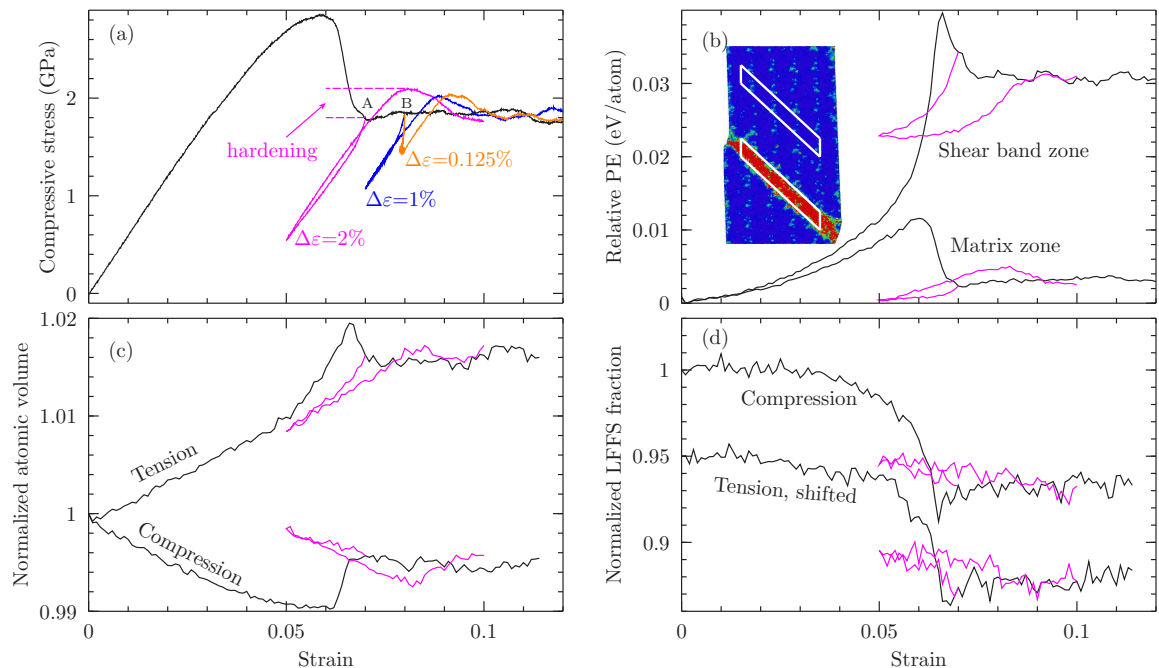}
\caption{(a) Engineering stress strain curves for a monotonic compressive loading (in black) and unloading/reloading (in color). The strain rate is 10$^8$/s except for the two unloading/reloading cases starting with point B. For the part $\varepsilon<0.08$ of these two cases, the rate is 5$\times10^7$/s for case $\Delta\varepsilon=$1\% and 6.25$\times10^6$/s for $\Delta\varepsilon$=0.125\%. (b) Average potential energy of two groups of atoms within the shear band region and the matrix region, as indicated in the inset, during the monotonic and cyclic loading. The inset picture corresponds to $\varepsilon=0.07$ during the monotonic loading. (c) and (d) Atomic volume and local fivefold symmetry (LFFS) in the shear band region during  monotonic and cyclic loading.}
\label{fig:strstr}
\end{figure*}

\section{Method}

The simulation details are as follows. First, a bulk cubic simulation cell containing 3375 atoms of each species was relaxed at 2000 K, about 700 K higher than the simulated melting point \cite{Tang2013}, for 2 ns. The liquefied system was then quenched to 100 K at 10$^{9}$ K/s. The deformation samples, around $49\times245\times490$ $ $\AA$^3$, were constructed from the final quenched configurations by repeating the latter $1\times5\times10$ times in $x$, $y$, and $z$ directions, respectively. The periodicity along $y$ was broken by inserting into the simulation cells a vacuum layer of about 500 $ $\AA$ $ to separate the samples from their periodic images. The samples with free surfaces were then relaxed at 600 K and 100 K, respectively, for 100 ps to reduce the structural periodicity caused by repeating the building blocks. Finally, a surface notch was created by removing some atoms in order to stimulate the formation of shear bands from the surface. The notched samples were then stretched or compressed along $z$ axis up to engineering strain $\varepsilon=0.2$ at a strain rate of 10$^8$/s. At some chosen $\varepsilon$ values, the state (i.e., the position and velocity of atoms) of the system was used to start a cyclic deformation by unloading the system from this state and then reloading the system. For the slide-stop-slide experiments, we chose a flow stress state ($\varepsilon=0.08$) and relaxed the system at this fixed strain for various times before further loading. All deformation tests were carried out at a temperature of 100 K.

All simulations were carried out under NPT (constant particle number, pressure, and temperature) ensemble using code LAMMPS \cite{lammps}, and some codes \cite{stukowski_visualization_2010,rycroft_voro_2009} were used for structure visualization and analysis. The atomic interactions were modelled by the embedded atom method (EAM) potential proposed for CuZr \cite{Mendelev2009}, the time step was set as 1 fs, and temperature and pressure were modulated with Nos\'{e}-Hoover thermostat and barostat, respectively. During the Voronoi polyhedra analyses, we assumed atomic radii of Cu and Zr to be 1.26 and 1.58 $ $\AA$ $ \cite{ward_structural_2013}, respectively, which well match the peaks of the radial distribution functions obtained in this work.

\section{Results and discussion}
\subsection{Cyclic loading process}

The stress strain curve of a typical compression test is shown in Fig. \ref{fig:strstr}(a). As can be seen, the stress starts to drop dramatically around $\varepsilon=0.06$, which is accompanied by the initiation and propagation of the shear band from the surface notch. The shear band passes through the sample and reaches its mature stage around $\varepsilon=0.07$ (see inset image  in Fig. \ref{fig:strstr}(b)), after which the flow stress becomes relatively stable. The stress strain curve of tension is similar to that of compression, but with the yield strength about 7\% lower. This difference was attributed to the pressure-dependence of yielding in metallic glasses\cite{schuh_atomistic_2003}. 

We begin the study of stress overshoot with the cyclic loading case. The overshoot effect is observed in both compression and tension, such that in the following we focus mainly on the compression case. We unloaded the system from $\varepsilon=0.07$ by $\Delta\varepsilon$=0.02 to $\varepsilon=0.05$ and an obvious stress overshoot after full reloading is observed, as shown in Fig.\ref{fig:strstr}(a). 

It is helpful to understand the overshoot from an energetic perspective. To this end we plot in Fig. \ref{fig:strstr}(b) the average potential energies of two groups of atoms, one along the shear band and the other within the matrix, as indicated in the inset image. During the monotonic loading case, the energy of the shear band group increases sharply upon the propagation of shear band and then decreases by some amount when the band passes through the sample and starts to slide, which releases some strain energy. In contrast, upon shear band propagation the energy of the matrix group drops due to the release of stress in the matrix. Upon cyclic loading the energy of the shear band zone decreases and becomes lower than that during the sliding stage of the monotonic loading case. From the energy perspective, this indicates that the shear band is stabilized during unloading and, hence becomes more resistant to shearing.

The resistance to atomic shearing has attracted enormous research interest and is widely believed to correlate with either free volume or structural symmetry \cite{cheng_indicators_2008}. A series of studies propose that atoms in the highly efficiently packed regions are more reluctant to shear than those in the regions with low packing efficiency \cite{wang_atomic_2015,zhong_deformation_2016}, while some studies argue that the symmetry order \cite{cheng_indicators_2008,cao_structural_2009,cheng_correlation_2009,peng_structural_2011}, especially the local fivefold symmetry (LFFS), is a better indicator of shearing resistance. Here we examine the evolution of both  atomic volume and LFFS during the cyclic loading process. For an atom, the atomic volume can be defined as the volume of its Voronoi polyhedron and the LFFS as the fraction of pentagon facets of the Voronoi polyhedron. While there are several different definitions for atomic free volume \cite{li_atomic_2009}, we believe atomic volume is a simple yet reasonable indicator for free volume when the volume of the atomic core is factored out.

As shown in Fig. \ref{fig:strstr}(c), the average atomic volume of the shear band region under monotonic compression reduces continuously by about 1\% and then rises abruptly upon shear band formation, followed by some minor fluctuation during the band sliding stage. It has been reported for metallic glass Cu$_{65}$Zr$_{35}$ that excess free volume is created after compression below the yield strength for several hours \cite{lee_structural_2008}. This is not inconsistent with our results since the excess free volume, resulting from permanent local plastic deformation, is measured after the compressive load being fully removed. For the unloading process the analysis of cyclic nanoindentation supposes that the free volume would decrease because the diffusion-controlled free volume annihilation process dominates over the shearing-controlled accumulation process \cite{yang_strain_2006}. Here, we observed that the change in volume during the unloading process depends on the stress state, with the volume decreasing and increasing for tension and compression, respectively. For both scenarios, the volume becomes slightly smaller than that in monotonic loading after the system is fully reloaded. This indicates that the free volume is overall reduced during the cyclic process although the change is small and comparable to the fluctuation in the plastic stage of monotonic loading.

\begin{figure}[t!]
\centering
\includegraphics[width=3.in]{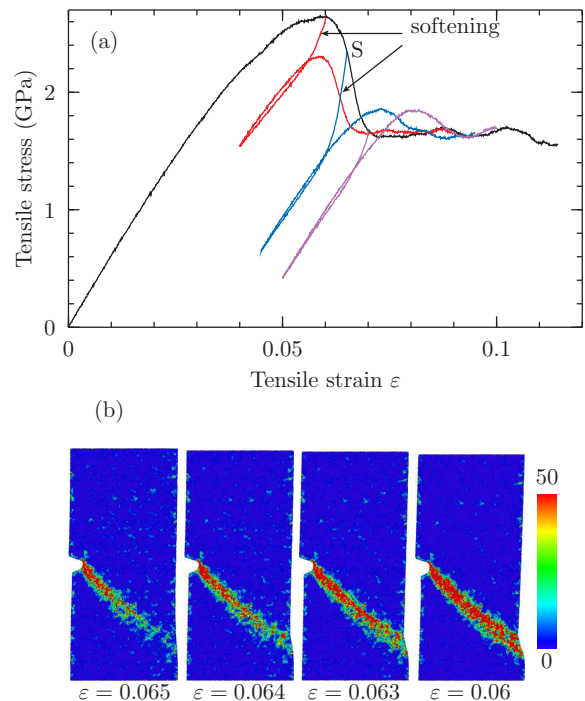}
\caption{(a) Stress undershoot or apparent softening during unloading from early or intermediate stage of shear band propagation. (b) Non-affine squared displacement ($ $\AA$^2 $) analysis indicates that the shear band continues to propagate when the sample is unloaded from $\varepsilon=0.065$ (point S in (a)).}
\label{fig:softening}
\end{figure}

\begin{figure*}[h!tb]
\centering
\includegraphics[width=6in]{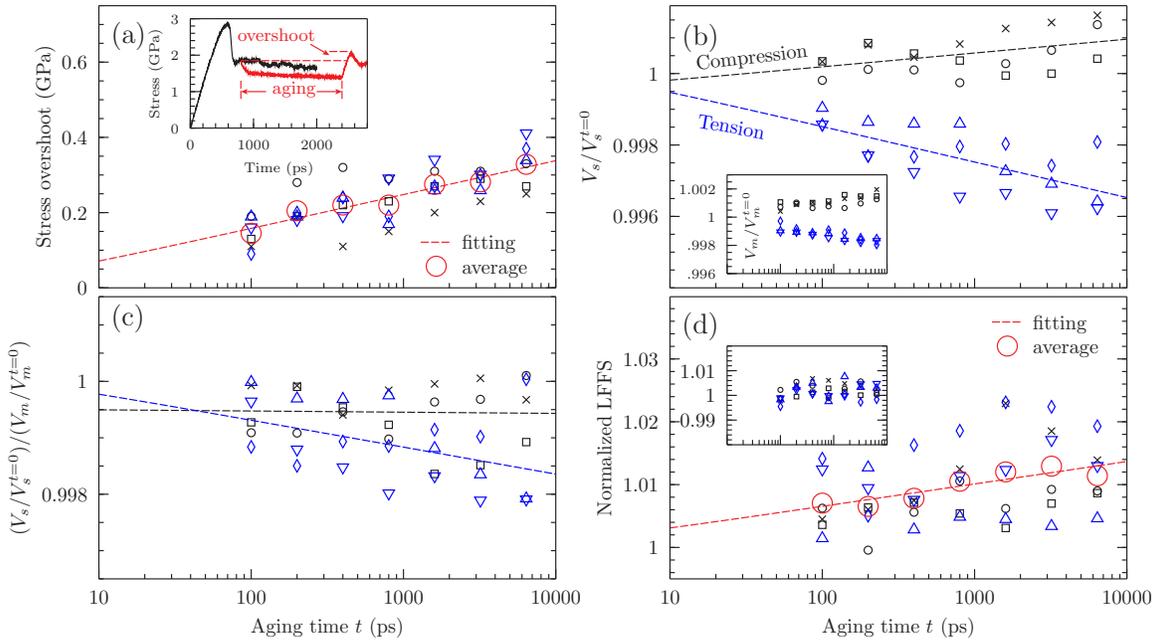}
\caption{(a) Stress overshoots with respect to aging time for three compression (black symbols) and three tension (blue symbols) tests. The fitting function is $f(t)=0.0392{\times\rm{log}}(t+1)-0.0225$. Inset: an example compression aging case. (b) Average atomic volume in the shear band zone ($V_s$), normalized by that at aging time $t$=0. Inset: Similarly normalized volume for the matrix. (c) The ratio of atomic volume in the shear band to that in the matrix, normalized by their values at aging time $t$=0.  (d) Average local fivefold symmetry (LFFS) in the shear band normalized to $t$=0. The fitting function is $f(t)=0.00155{\times\rm{log}}(t+1)+0.999$. Inset: Similarly normalized LFFS for the matrix. }
\label{fig:aging}
\end{figure*}

The average LFFS of the shear band region remains constant during the elastic deformation stage, sharply decreases during the shear band propagation, and then fluctuates slightly around 93\% of its original value during shear band sliding, as seen in Fig. \ref{fig:strstr}(d). Upon unloading the LFFS increases, for both compression and tension, because of structural relaxation and after the system is fully reloaded, the fivefold symmetry  is about 1-1.5\% higher than that during the monotonic loading and then gradually merges with the latter. 

Although the stress overshoot upon reloading is obvious, it is interesting to note that the elastic moduli associated with monotonic loading and reloading are nearly identical, as shown in Fig. \ref{fig:strstr} and Fig. \ref{fig:softening}. The flow stress upon reloading depends only on the band because the shear band region has a lower strength than the matrix. However, modulus explicitly depends on the strain upon certain stress. This implies that the elastic modulus of the sheared sample depends not only on the modulus of the shear band, but also of the matrix. Conceptually, the sample can be viewed as a serial combination of two springs characterized by the shear band region and the matrix region, respectively. In view of the relatively small structural change in the shear band and, more importantly, the small volume fraction of the band, it is not surprising that the elastic modulus during reloading is similar to that of monotonic loading.

By fixing the unloading/reloading time, we tested the influence of strain rate on the overshoot. We chose the unloading/reloading rate to be 0.5, 0.25, 0.125, and 0.0625 times the original strain rate 10$^8$/s, which correspond to unloaded strain $\Delta\varepsilon$=1\%, 0.5\%, 0.25\%, and 0.125\%, respectively, and after full reloading the strain rate was restored to 10$^8$/s. It is found that the overshoot amount is relatively insensitive to the strain rates when the relaxation time is fixed, as shown in Fig. \ref{fig:strstr}(a). 

Before moving to the discussion of the slide-stop-slide process, we note that the unloading-reloading behaviour depends on the shear band state at the start of unloading. When unloading starts from a mature shear band, usually a nearly elastic unloading-reloading is observed, followed by the stress overshoot. On the other hand, if the system is unloaded from the early or intermediate stage of shear band propagation, apparent stress undershoot or softening occurs before the elastic unloading process starts, and the consequent overshoot may or may not occur. This is illustrated by the two unloading-reloading curves starting from $\varepsilon=0.06$ and 0.065 in Fig. \ref{fig:softening}(a). The unloading softening results from the fact that, upon unloading, the shear band continues to propagate for some time, as shown in Fig. \ref{fig:softening}(b). The propagation of the shear bands even upon unloading provides a possible mechanism of degradation or fatigue for metallic glasses since in some metallic glasses \cite{menzel_stress-life_2006,menzel_fatigue_2006} fatigue damage can initiate in the form of shear bands at stresses as low as 10\% of the ultimate tensile strength.

\subsection{Slide-stop-slide process}
To explicitly simulate the slide-stop-slide process, we held the samples at $\varepsilon=0.08$ for various times before resuming the loading. As shown in the inset of Fig. \ref{fig:aging}(a), reloading after the relaxation causes a stress overshoot. We tested three compression and three tension samples and the results indicate that, statistically, the overshoot increases logarithmically with respect to the aging time, as shown in Fig. \ref{fig:aging}(a), which is qualitatively consistent with experimental findings \cite{maass_shear-band_2012}. By analyzing non-affine squared displacement \cite{falk_dynamics_1998} we found that structural relaxation accompanying the aging process is mainly in the shear band zone, whereas it is negligible in the matrix. As shown in Fig. \ref{fig:aging}(b), for both matrix and shear band, the average atomic volume increases during the compressive aging but decreases during the tensile aging. This implies that a large part of the volume change within the shear band could be only due to the elastic effect as seen in the matrix. We also note that the slopes of the fitting lines for the volume change are different for compression and tension, which is consistent with the asymmetricity of interatomic force around the equilibrium distance. To factor out the elastic effect and evaluate the change in free volume, we plot in Fig. \ref{fig:aging}(c) the ratio of atomic volume change within the shear band to that within the matrix. As can be seen, for the tension case, the atomic volume within the band decreases more rapidly than that in the matrix and, thus, an annihilation of free volume occurs within the band during aging. However, for compression almost no free volume annihilation occurs within the shear band and all the change in atomic volume can be attributed to the elastic recovery effect. On the other hand, the LFFS of the shear band increases in both compression and tension cases (Fig. \ref{fig:aging}(d)), while that of the matrix is nearly constant. Like the stress overshoot, the change in volume and LFFS can be fitted logarithmically.
 
\begin{figure}[tb]
\centering
\includegraphics[width=3.in]{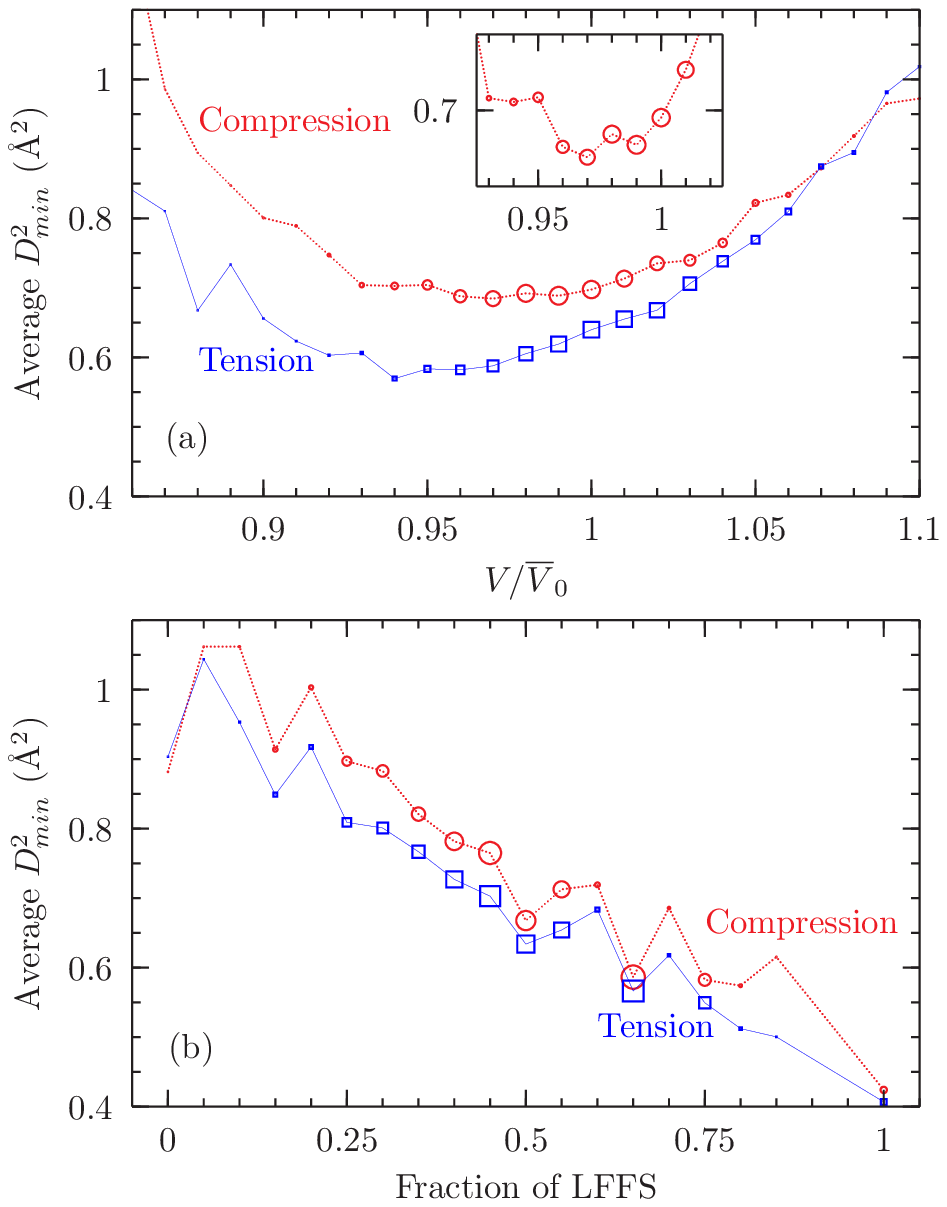}
\caption{The relationship between non-affine squared displacement and normalized atomic volume (a) and local fivefold symmetry (LFFS) (b). The radial size of the symbols represents their relative probability.  Both shear band and matrix atoms were used for data collection, and the volume and LFFS data were averaged for  four configurations of a monotonic compression test at strains of 0.07, 0.08, 0.09, and 0.10, and the displacements are relative to those 10 ps later, correspondingly. $V$ is atomic volume and $\overline{V}_0$ is the average atomic volume (species dependent) of the quenched bulk system at 100 K.}
\label{fig:vol-5fold}
\end{figure}

Fig. \ref{fig:vol-5fold} shows the correlation between the minimum non-affine squared displacement $D^2_{min}$, which is a good indicator of shearing resistance, and atomic volume and LFFS. The parabolic displacement-volume curves indicate that atoms tend to shear when their volume deviate from the equilibrium, which is about 3\% smaller than the as-quenched average volume $\overline{V}_0$ in this work. A similar parabolic relationship was found in our previous work \cite{tang_formation_2016} and in other systems \cite{yang_structures_2016}. For the major atoms their volume is nearby the bottom of the parabola curves where shearing resistance is relatively insensitive to volume change, although the correlation is a bit more pronounced for the tension case. Compared with atomic volume, LFFS exhibits a roughly linear relationship with $D^2_{min}$. We conclude that the relaxed local structural symmetry is probably responsible for the stress overshoot. Interestingly, the atoms with high (low) LFFS tend to decrease (increase) their symmetry during the aging process, which is similar to the rearrangement of the `+' and `$-$' states in the STZ model mentioned earlier.

\begin{figure}[tb]
\centering
\includegraphics[width=3.3in]{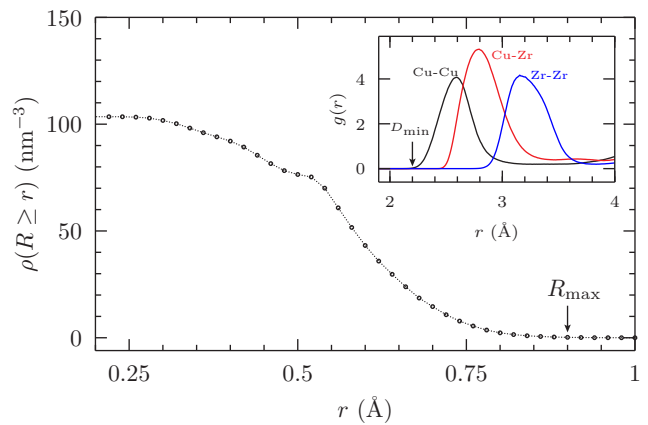}
\caption{The cumulative density $\rho$ distribution of subatomic vacancies as a function of vacancy radius $R$ within the relaxed undeformed bulk metallic glass at 100 K. Atomic radii are defined as 1.26 and 1.58 $ $\AA$ $ for Cu and Zr, respectively, determined from  the radial distribution function $g$($r$). $R_{\rm{max}}$ is the approximate maximum vacancy radius and $D_{\rm{min}}$ is the possible minimum diameter of a vacancy that allows an atom to hop into.}
\label{fig:vac}
\end{figure}

We also analyzed the distribution of atomic vacancies or voids within the samples using an existing algorithm \cite{Levchenko2011a}. Being close to the original meaning of `free volume' proposed by Spaepen \cite{spaepen_microscopic_1977}, a vacancy is defined as spherical excess space which does not overlap with an atom or another vacancy. As shown in Fig. \ref{fig:vac}, only subatomic vacancies (with radius smaller than $\sim$0.9 $ $\AA$ $) exist within the system, and, according to the radial distribution function, within the available energy fluctuations an atom can only hop into a vacancy site with radius larger than $\sim$1.1 $ $\AA$ $. Clearly, even upon severe deformation it is unlikely or extremely difficult to produce any vacancies of atomic size that could significantly affect the mechanical or diffusional properties. Indeed, the pressure independence of diffusivity indicates that atomic diffusion in metallic glasses is not via atomic jumps \cite{faupel_diffusion_2003, wang_source_2015}.

With the connection between fivefold symmetry and the stress overshoot established, it is important to understand the underlying mechanism. Fundamentally, the high resistance of clusters exhibiting fivefold symmetry to shear can be linked to their energetic stability. From the thermodynamic perspective, computations of isolated Zr-Pt clusters have shown that an ideal 13-atom icosahedron cluster can be lower in potential energy by $\sim$10 eV than an ideal face-centered-cubic cluster, although the distorted icosahedra, which appear in real systems, are somewhat higher in energy \cite{hirata_geometric_2013}. Our previous computations of CuZr relaxations at various quench rates also indicate that the systems with lower energy, which are more slowly quenched, also contain a higher fraction of icosahedra \cite{tang_formation_2016}. Moreover, icosahedra can inter-penetrate each other to form medium range order and further enhance their stability \cite{wakeda_icosahedral_2010,ding_full_2014}. From a kinetic perspective, it has been shown that atoms in clusters of high fivefold symmetry have low mobility \cite{hu_five-fold_2015}, which implies a high saddle point or energy barrier for the breakdown of these clusters. With the above said, we note that in systems with low icosahedra fraction the mechanical behaviour may be controlled by some other topologically stable cluster, which is worth further investigation, although it has been shown for various systems\cite{hu_five-fold_2015} that the LFFS fraction substantially increases upon glass transition.

\subsection{Monotonic deformation process}
\begin{figure}[htb]
\centering
\includegraphics[width=3.3in]{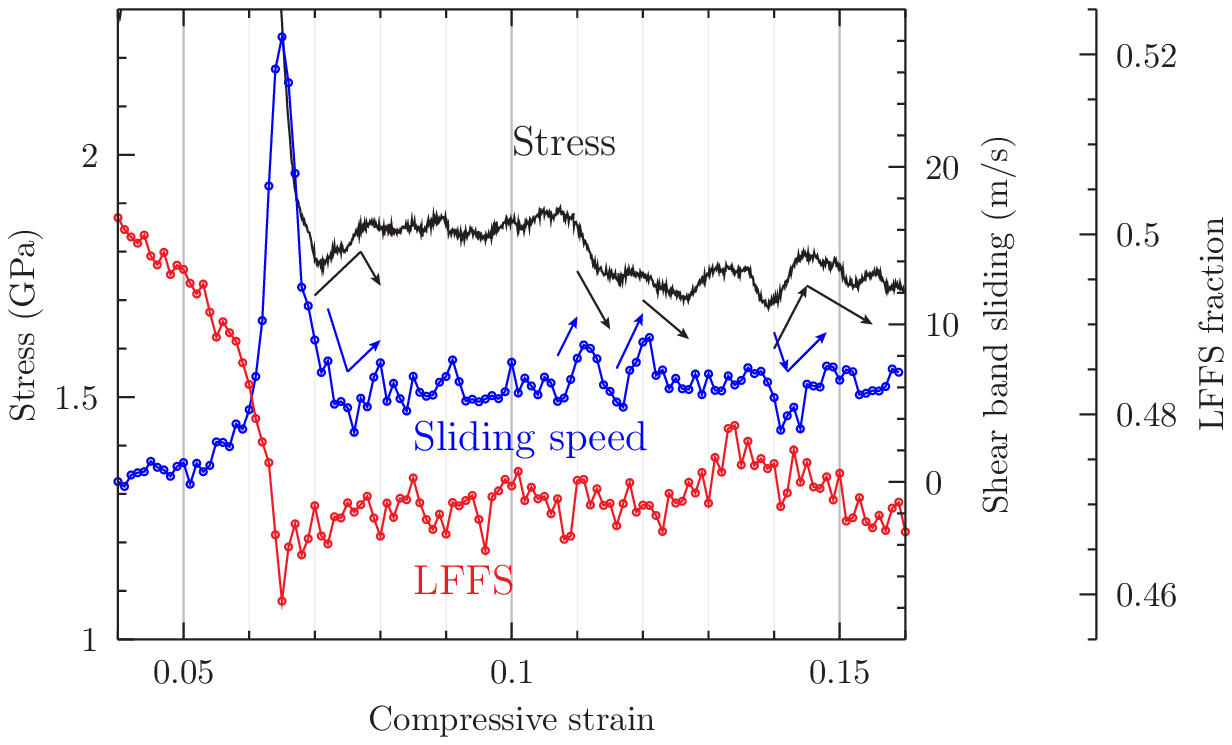}
\caption{The plastic stage of the compressive stress strain curve that is shown in Fig. \ref{fig:strstr}(a), the sliding speed of the corresponding shear band, and the fraction of local fivefold symmetry (LFFS) in the shear band.}
\label{fig:slide}
\end{figure}

Finally, we discuss the overshoot in the monotonic loading case or the serration of flow stress. A close examination of the stress strain curves shows that the flow stress exhibits some fluctuations, as shown in Fig. \ref{fig:slide}, somewhat resembling the serrations in compression experiments. Such fluctuations have also been observed in previous simulations of Cu-Zr system \cite{cao_structural_2009} and Lennard-Jones systems \cite{cao_nanomechanics_2018}.   By checking the relative displacement of two reference atoms on either side of the shear band, we found the sliding speed of the shear band also fluctuates. Nevertheless, we note that the fluctuations observed here are different from the stress serrations observed exmperimentally because the sliding speed in this study is always larger than zero, while experimentally the shear band does not slide during the stress rise phase of serrations \cite{song_capturing_2010,maass_shear-band_2015}. 

The absence of stress serrations in our simulations is not surprising in view that, experimentally, serrations occur at low strain rates (usually around 10$^{-3}$/s) and our simulated strain rates are about 11 orders of magnitude higher. We also tried strain rate of 10$^6$/s but the results are qualitatively not too different. We noted that LFFS in the shear band region also fluctuates (Fig. \ref{fig:slide}), but statistically the fluctuation exhibits a poor match with the stress fluctuation. This indicates that the effect of symmetry relaxation is difficult to accumulate during a fast dynamic process simply because the externally applied strain overwhelms the relaxation effect. However, in a slow enough loading process we expect that the accumulated relaxation of local symmetry would result in overshoot and, hence, flow stress serrations. 

\begin{figure}[htb]
\centering
\includegraphics[width=3.in]{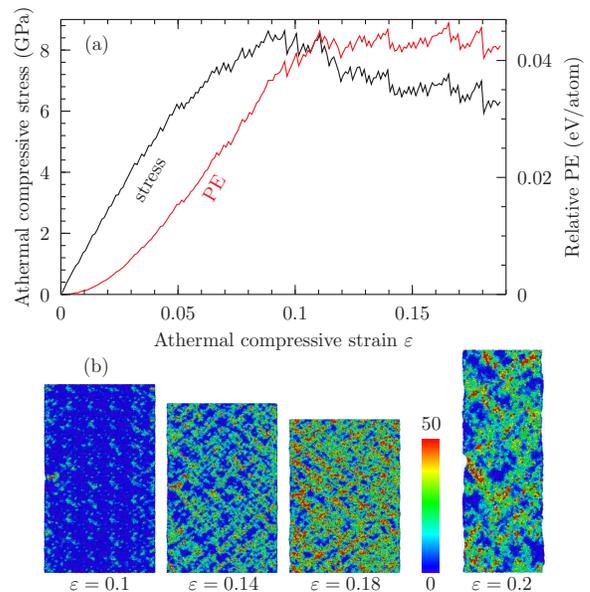}
\caption{(a) Stress and relative potential energy (PE) as a function of strain during athermal quasistatic compression. During the compression, incremental strain 0.001 followed by geometry optimization was applied. (b) Non-affine squared displacement ($ $\AA$^2 $) distribution in the sample at different strains. The image of $\varepsilon=0.2$ is for tension of an as-cast monolithic sample.}
\label{fig:athermal}
\end{figure}

In view of the difficulty for molecular dynamics simulations to reach the experimental timescales or strain rates, some authors \cite{maeda_atomistic_1981} introduced the athermal quasistatic (AQS) algorithm that consists of iterative application of a small strain to the system, whereby each strain application is followed by a minimization of the potential energy. The AQS algorithm is based on the assumption that during quasistatic deformation the system remains close to a mechanically stable state, which can be approximated by a local minimum in the potential energy landscape. For completeness, we also performed AQS compression simulations by applying an incremental strain of 0.001 to the system and then minimizing the potential energy of the whole system. As can be seen from Fig. \ref{fig:athermal}(a), the flow stress serrations, accompanied by potential energy serrations, are sharper and more regular compared with the molecular dynamics case and qualitatively similar to those observed in the Lennard-Jones systems \cite{cao_nanomechanics_2018}. Nevertheless, unlike molecular dynamics, the AQS algorithm does not include the time factor and so the information about deformation speed during stress serrations is not available.

Further, we note that the serrations observed in the AQS simulations are different to the serrations caused by the highly localized shear bands. In AQS simulations, the increase in stress corresponds to the climb, driven by the applied strain, of the system in the potential energy landscape around a local minimum, and the drop indicates that the system finds a new local minimum after passing a stability limit \cite{maloney_amorphous_2006}. Because the energy minimization is applied to the whole system, a drop in stress means a global avalanche of shear events, instead of highly localized events as in shear bands, as shown in Fig. \ref{fig:athermal}(b). These two different modes of shear events probably are the reason for the very different yield stresses observed in the AQS and molecular dynamics simulations. 

We noted that the way in which we built the samples, i.e., periodically combining the building block into a larger sample, may affect the mechanical response of the samples. This is manifested by the periodic distribution of the atoms with higher non-affine squared displacement ($\varepsilon=0.1$ in Fig. \ref{fig:athermal}(b)), although this effect is somewhat hidden in the cases of molecular dynamics deformation. To validate our observations, we performed AQS tension on an as-cast monolithic sample (${\sim}26{\times}130{\times}260~$\AA$^3$) and similar global shear-banding was observed ($\varepsilon=0.2$ in Fig. \ref{fig:athermal}(b)).

\section{Summary}

We have studied the stress overshoot of metallic glass Cu$_{50}$Zr$_{50}$ within the inhomogeneous deformation regime based on molecular dynamics simulations of the cyclic loading process and the slide-stop-slide process. The simulations were performed in both uniaxial tensile and compressive loading and similar overshoot effects were observed. For cyclic loading, stress overshoot is observed upon full reloading if the shear band is already mature before the start of unloading. On the other hand, if the sample is unloaded from the propagation stage of shear banding, an apparent stress undershoot occurs. For the slide-stop-slide process, stress overshoot is observed after the loading is resumed and the overshoot increases logarithmically with respect to the stop time, consistent with the experiments.

It was found that, due to elastic recovery, atomic volume in the shear band region increases for the compression case but decreases for the tension case. After factoring out the elastic effect, the atomic volume (indicative of free volume) remains essentially unchanged for compression but decreases for tension. On the other hand, local fivefold symmetry consistently increases for both scenarios. These findings indicate that local symmetry order, instead of free volume, is responsible for the stress overshoot. 

While we did not directly observe flow stress serrations in monotonic loading cases due to the inaccessible low strain rates in atomistic simulations, symmetry relaxation provides a reasonable atomistic mechanism for the flow stress serrations. On the other hand, athermal quasistatic simulations produce stress serrations, but such serrations result from a global avalanche of shear events instead of the relaxation of shear bands. %\hlt{Overall, our studies reveal that atomistic-scale symmetry order impacts on the dynamics of shear-banding and, hence, mechanical behaviour of metallic glasses at low temperatures.}

\section*{Acknowledgements}
The authors acknowledge NCI National Facility for computational support of project codes eu7 and y88. Chunguang Tang would particularly like to thank the Australian Research Council for the DECRA Fellowship (grant no. DE150100738) for enabling this work to be carried out.
%%%%%%%%%%%%%%%%%%%%%%%%%%%%%%%%%%%%%%%%%%%
% ********* main text ends here *************
\section*{Reference}
%\bibliography{zotero,reference,tang}
%\bibliographystyle{./mystyle-acta.bst}

\end{document}